\documentclass[11pt,twocolumn,letterpaper]{article}

\usepackage[utf8]{inputenc}
\usepackage[T1]{fontenc}
\usepackage{float}
\usepackage{adjustbox}
\usepackage[a4paper,top=2.5cm,bottom=2.5cm,left=2cm,right=2cm]{geometry}

\usepackage{amsmath}
\usepackage{amssymb}
\usepackage{amsthm}
\usepackage{mathtools}
\usepackage{listings}
\usepackage{enumitem}[shortlabels]
\usepackage[colorinlistoftodos]{todonotes}
\usepackage[colorlinks=true, allcolors=blue]{hyperref}
\usepackage{url}

\usepackage{tcolorbox}

\usepackage{authblk}

\title{
		\vspace{-0.7in} 	
		\usefont{OT1}{bch}{b}{n}
		
    \begin{minipage}{10.7cm}
    	\begin{center}

\normalfont \normalsize \textsc{submitted to Frontiers} \\
		\huge Efficient High Performance Computing with the \mbox{ALICE}  Event Processing Nodes GPU-based farm\\
        
        \normalsize \rmfamily \mdseries
        \vspace{0.1in}
Federico~Ronchetti\,$^{*1,5}$, Valentina~Akishina\,$^{2}$, Edvard~Andreassen\,$^{1}$, Nora~Bluhme\,$^{2}$, 
Gautam~Dange\,$^{2}$, Jan~de~Cuveland\,$^{2}$,  Giada~Erba\,$^{1}$, Hari~Gaur\,$^{2}$, Dirk~Hutter\,$^{2}$, Grigory~Kozlov\,$^{4}$, Luboš~Krčál \,$^{1}$, Sarah~La~Pointe\,$^{2}$, Johannes~Lehrbach\,$^{2}$, Volker~Lindenstruth\,$^{2,3,4}$, Gvozden~Neskovic\,$^{2}$, Andreas~Redelbach\,$^{2}$, David~Rohr\,$^{1}$, Felix~Weiglhofer\,$^{2}$, Alexander~Wilhelmi\,$^{2}$\\
        \vspace{0.1in}
        \small
		\mbox{$^{1}$~European~Organization~for~Nuclear~Research~(CERN),~Geneva,~Switzerland} \\
		\mbox{$^{2}$~Frankfurt~Institute~for~Adv5anced~Studies,~Frankfurt~am~Main,~Germany}  \\
		\mbox{$^{3}$~Goethe~University,~Frankfurt~am~Main,~Germany} \\
		\mbox{$^{4}$~GSI~Helmholtz~Centre,~Darmstadt,~Germany} \\ 
		\mbox{$^{5}$~INFN,~Laboratori~Nazionali~di~Frascati,~Frascati,~Italy} \\ 
		\mbox{$^{*}$~Corresponing author} \\

        \normalsize
        \date{\date{\vspace{0.1in}\today}}
    	\end{center}
     
    \end{minipage}\hfill

\author{}

\vspace{0.1in}
\begin{tcolorbox}[colframe=white, arc=8pt]
\normalsize \rmfamily \mdseries
\centering
\textbf{Abstract}\\
 Due to the increase of data volumes expected for the LHC \mbox{Run 3} and Run~4, the \mbox{ALICE} Collaboration designed and deployed a new, energy efficient, computing model to run Online and Offline (O$^2$) data processing within a single software framework. The \mbox{ALICE} O$^2$ Event Processing Nodes (EPN) project performs online data reconstruction using GPUs (Graphic Processing Units) instead of CPUs and applies an efficient, entropy-based, online data compression to cope with Pb–Pb collision data at a 50~kHz hadronic interaction rate. Also, the O$^2$ EPN farm infrastructure features an energy efficient, environmentally friendly, adiabatic cooling system which allows for operational and capital cost savings.  \\
\textit{Keywords:} \textbf{scientific computing, sustainable computing, HTC, HPC, GPU}
\end{tcolorbox}
\vspace{-0.7in} 
}

\hypersetup{
    colorlinks=true,
    linkcolor=blue,
    filecolor=magenta,      
    urlcolor=cyan
    }

\begin{document}

\maketitle

\section{Introduction}

The Large Hadron Collider (LHC) accelerator at CERN returned to full operation on July 5$^{th}$, 2022 when proton–proton collisions occurred at a record center-of-mass energy of 13.6 TeV and data taking activities resumed. During the LHC shutdown (2019–2021), the \mbox{ALICE} detector \cite{alice-det} underwent a substantial upgrade \cite{alice-up} providing improved track reconstruction, and an increased interaction rate of up to 50 kHz for Pb–Pb collisions in continuous readout mode. These improvements enabled recording a data sample a factor of 10 larger than the combined Run 1 and Run 2 samples.

\section{The upgraded \mbox{ALICE} detector}
The \mbox{ALICE} detector comprises a central barrel (by far the largest data producer in the system) and a forward muon arm. The central barrel relies mainly on four sub-detectors for particle tracking:
\begin{enumerate}
    \item the new Inner Tracking System (ITS) which is a 7-layer, 12.5 gigapixel monolithic silicon tracker 
    \item the upgraded Time Projection Chamber (TPC) with GEM-based readout for continuous operation
    \item the Transition Radiation Detector (TRD)
    \item the Time-Of-Flight (TOF)
\end{enumerate}

 The muon arm is composed of three tracking devices:
 \begin{enumerate}
     \item a newly installed Muon Forward Tracker (a silicon tracker based on the same monolithic active pixel sensors used for the new ITS)
     \item a revamped Muon Chambers system
     \item a Muon Identifier (previously a trigger detector adapted to run in continuous readout).
 \end{enumerate}
 
\subsection{The O$^2$ EPN project}
The major ALICE hardware upgrades for Run 3 necessitated the development and implementation of a completely new computing model: the O$^2$ project. This project unifies online (synchronous) and offline (asynchronous) data processing into a single software framework, enabling the same code-base to be executed in both contexts with appropriate selections and parameters. Naturally, the O$^2$ computing model necessitated a comprehensive overhaul and significant scaling up of the experiment's computing farms to accommodate the demands of data readout and processing. To address these challenges, the \mbox{ALICE} EPN farm consists of 350 high-performance servers, each equipped with four boards for a total of 2800 GPUs.

With compressed data rates already reaching 1--2~PB/day, the dramatic increase in data volumes compared to Run~2 made storing raw data entirely impractical, driving the need for efficient online compression and the adoption of GPUs in place of CPUs to accelerate reconstruction tasks. GPUs, with their superior compute throughput enabled by intrinsic parallelism, offer both cost and energy savings compared to CPU-based solutions. Without GPUs, approximately eight times as many CPU-only servers of the same type, along with additional resources, would be required for the online processing of TPC data from Pb–Pb collisions at a 50 kHz interaction rate (corresponding to an instantaneous LHC luminosity of \mbox{$6\times10^{27}$~cm$^{-2}$s$^{-1}$}) \cite{chep23}.

\section{\mbox{ALICE} Computing for \mbox{Run 3} and beyond}
The ALICE Collaboration has been a pioneer in leveraging GPUs for data compression and online processing since 2010, starting with the High-Level Trigger (HLT) computing farm \cite{hlt}. The HLT had direct access to detector readout hardware and raw data, playing a critical role in Run~2 by enabling data compression for heavy-ion collisions. Its software framework was already advanced enough to perform online data reconstruction using GPUs. The operational experience gained with the HLT farm during Runs~1 and Run~2 was instrumental in shaping the design and development of the current O$^2$ software and hardware systems.

\subsection{Computing infrastructure}\label{computing-infra}
\begin{figure}[h!]
    \centering
    \includegraphics[width=0.75\linewidth]{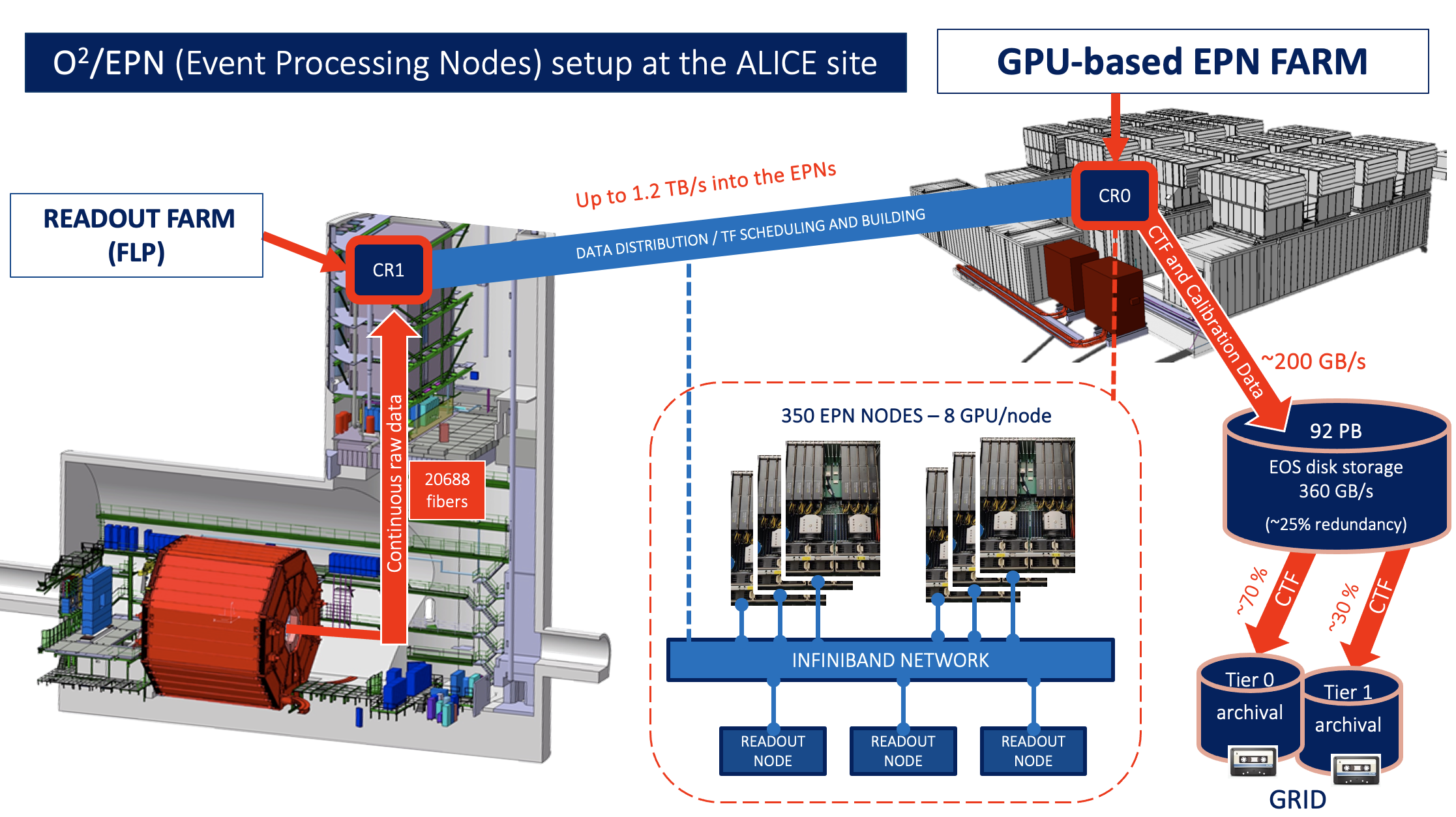}
    \caption{\small Schematic overview of the O$^2$ data~flow and the \mbox{ALICE} computing infrastructure. See Section \ref{computing-infra}. Additional information in \cite{courier}}
    \label{fig:dataflow}
\end{figure}

Figure~\ref{fig:dataflow} illustrates the ALICE data readout and processing workflow in Run~3. The detectors' front-end electronics boards are connected to custom Field-Programmable Gate Array (FPGA) boards, capable of continuous readout and Zero Suppression (ZS), which are hosted on the First Level Processor (FLP) farm nodes located near the experimental cavern.
The connecting links are based on the GBT (GigaBit Transceiver) architecture, a versatile and high-performance communication framework developed by CERN for use in high-energy physics experiments. GBT is designed to handle the stringent data transmission requirements of particle detectors, providing robust, low-latency, and high-throughput (4.8~Gb/s) links between the detector front-end and readout electronics. 
The TPC outputs a prohibitive raw data rate of 3.3 TB/s at the front end, so ZS in the FPGA is crucial for data rate reduction, providing manageable rates, especially for high interaction rate Pb—Pb collisions. The FLP and EPN farms handle the high data throughput using an InfiniBand (IB) network fabric and a custom software suite (detailed later in the text). The compressed, reconstructed data produced by online processing are transferred to the central CERN Information Technology (IT) data center, located a few kilometers from the ALICE experimental site, via dedicated links employing standard TCP/IP over Ethernet. 
A seamless and balanced exchange between the EPN farm's IB fabric and CERN IT is facilitated by four IB-to-Ethernet gateways positioned near the EPN farm’s five core 200 Gb/s IB managed switches. The CERN data center provides approximately 150 PB of storage, managed by sophisticated storage policies (see Section \ref{ctf}) to ensure efficient use of buffer space during different phases of experiment data taking. 

\subsection{The \mbox{ALICE} O$^2$ EPN farm}  
The EPN farm data~center was built on the surface at the LHC Point~2 experimental site, was designed using a modular approach (hence it is located farther from the ALICE cavern than the FLP farm). 
The concept was implemented using IT containers so that the data center could easily scale to meet any potential need for additional IT equipment. 

The ALICE data center comprises four containers designed to house IT equipment. Currently, three of these containers are used to host the EPN farm worker nodes, infrastructure nodes, and network equipment, while the fourth serves as a utility container for the time being.  The layout is depicted in the top-right corner of Fig.~\ref{fig:dataflow}. In addition to the IT containers, the figure also shows the two service containers that supply the necessary power and purified water to operate and cool the IT infrastructure.

As for any modern high-performance computing (HPC) workload, hardware and software co-design is an important factor of the design phase in order to reduce costs and manage environmental resources appropriately. The compute requirements for \mbox{ALICE} are not expected to change drastically during the lifetime of the experiment, so the data center was tailored for the needs of Run~3 and 4, including the upcoming EPN farm refurbishment foreseen for the LHC Long~Shutdown~3 (2027-2029). 

\subsection{Technical choices}\label{tech-choice}
The IT containers' operating environment (temperature and humidity) is regulated using airflow managed by a modular adiabatic cooling system, consisting of four independent air handling units (AHUs) per container. These AHUs are powered by a single power line, with an automated transfer switch (ATS) automatically switching them to a secondary line in case of a primary line failure. This transition results in a brief interruption of cooling, as the process effectively mimics a power outage. All AHUs are connected to a single ATS, meaning they operate on the same power line and will collectively shut down if the primary line fails.

During routine maintenance, individual AHUs are taken offline for inspection, cleaning, or air filter replacement. However, farm operations remain unaffected, as the remaining AHUs maintain adequate cooling during these interventions. If one AHU in a container goes offline, the remaining three activate a "boost mode" to compensate for the reduced ventilation capacity.

Adiabatic cooling has become an increasingly popular choice for IT data-centers as it offers several advantages over traditional cooling such as better energy efficiency and reduced power usage. Adiabatic cooling is significantly more energy-efficient compared to conventional cooling as it exploits the process of evaporation to cool air, which requires less electricity than mechanical refrigeration. Also, unlike traditional chillers and compressors, adiabatic cooling systems use fans and water, which consume less power so data centers using adiabatic cooling can achieve a lower power usage effectiveness (PUE) rating, indicating higher energy efficiency.

The PUE \cite{pue} can be defined as the ratio between the total power consumed by the IT facility ( including cooling, water purification, etc.) over the IT-related power. Keeping the PUE close to 1 improves the overall data-center energy efficiency, directly impacting on the operational and capital cost savings since lower electric energy consumption reduces cost and maintenance needs.
In terms of environmental benefits, adiabatic cooling can be considered environmentally friendly with respect to mechanical cooling as its lower energy consumption means less greenhouse gas emissions and less water consumption. In a real-world scenario the PUE can deviate temporarily from one depending on the amount of IT load. 

The EPN farm’s total power consumption peaks at 550 kW at the start of a \mbox{Pb--Pb} physics fill, when the hadronic interaction rate reaches 50 kHz. During this period, cooling power typically averages around 26~kW. Since \mbox{Pb--Pb} collisions occur in cooler months, adiabatic cooling is not required, and the AHUs operate using air-to-air heat exchange only. Under these conditions, the farm achieves an PUE of 1.05.

PUE values increase slightly under different operational scenarios. For instance, during proton collisions at lower rates (650 kHz), which typically occur in spring/summer, hence with the adiabatic cooling operational, the PUE can rise to a maximum of 1.16.

\begin{figure}
    \centering
    \includegraphics[width=0.65\linewidth]{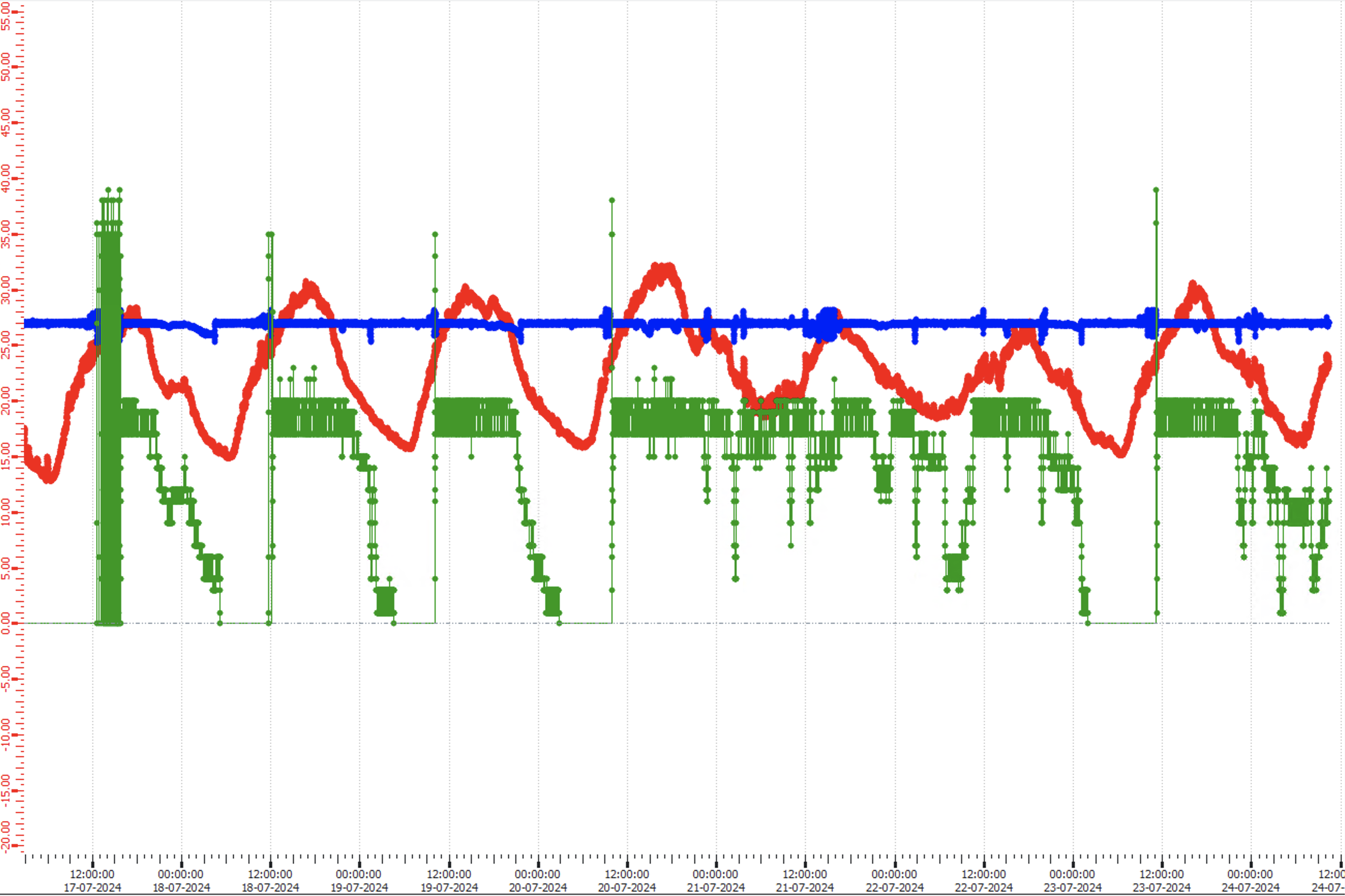}
    \caption{\small Correlation between the adiabatic mode of one Air Handling Unit (AHU) and the outside ambient temperature (red markers). Each time the supply air temperature (blue markers) exceeds the value $T_{start}$ the pumps are activated (green markers, expressing the pump speed in opening grade from 0 to 100) and the supply air to the racks is cooled down until the value $T_{start} -H$ is reached, where $T_{start} = T_{set} + \Delta T$ with $T_{set} = 27$~C and   $\Delta T =1$~C. The quantity $H$, also expressed in degrees C, represents the pumps hysteresis and is related to the head load generated in the container: for the EPN adiabatic cooling system $H = 2.2$~C. See Section \ref{tech-choice}. }
    \label{fig:temp-adiabatic}
\end{figure}

The ALICE EPN IT infrastructure operational range allows a good cooling performance for different climatic conditions. At the \mbox{ALICE} geographical location, the pumps run water and a nozzle system sprays aerosol on the heat exchangers only when the outside air temperature is higher then  $T_{start}$  (adiabatic system in "summer" mode, see Fig.~\ref{fig:temp-adiabatic}). Roughly speaking, when the ambient temperature is few degrees below the set-point the cooling units exchange heat with the outside air without the need of vaporizing water. Adiabatic cooling also improves the data-center air quality by using  air filter sets and by fully decoupling the internal and external air circuits. In addition, humidity is controlled to help prevent electrostatic discharge. 

One of the \mbox{ALICE} technical requirements was a cooling capacity of up to 1~kW per rack height unit \cite{o2-tdr19} to allow for servers with as many GPUs as possible. However, the average cooling power per rack in the final setup turned out to be significantly below 1~kW, only around 600~W/U (where U=1.75~inches represents the standard unit of measurement for the height of equipment designed to fit in an IT rack). However, given the available rack space and the good peak cooling power per rack height unit, the required flexibility is granted by the chosen setup.
The difference between average and peak cooling capacity was validated with intense testing, to ensure the cooling system could reliably remove up to 1~kW/U, without creating any hot spots. 

The ALICE IT infrastructure is designed to accommodate denser server units in the future. Ongoing studies aim to determine the most suitable hardware accelerator technology for upgrading the EPN farm during the LHC's Long~Shutdown~3. In Run 4, data rates are expected to increase by an additional $\sim$20\%, driven by the upgrade of ALICE's inner layers of the ITS pixel tracker and the installation of a new electromagnetic calorimeter. Furthermore, the consolidation of the FLP farm will result in higher data throughput to the EPNs, potentially reaching a theoretical maximum of 600 Gb/s per InfiniBand (IB) link. Given current market trends in HPC GPU computing, it is anticipated that the upgraded farm could fit within a single container. This setup would allow the operation of two farms in parallel: the new hardware could handle synchronous and prompt asynchronous processing, while the existing farm could be repurposed as a GPU-enabled LHC world wide computing grid (GRID) node.

\subsection{IT installation}
The first container units were delivered in late 2018, and the complete data center was finalized at the end of 2019, becoming fully operational. The key milestones for the installation and commissioning of the O$^2$ system are detailed in \cite{o2-evol}, with the first prototype system available as early as July 2019. The first \mbox{Run 3} servers, equipped with final hardware configurations, were shipped at the end of 2020 and became operational in early 2021.

The EPN farm employs taller 48U racks, which provide the advantage of closer proximity between servers and the top-of-rack (TOR) switches, enabling more efficient network integration. High-speed networks operating at speeds exceeding \mbox{100~Gb/s} are constrained by the limited range of passive direct attached copper (DAC) cables \cite{nvidia-dac}. These limitations arise due to the high-frequency electrical signals, which experience increased losses, such as higher insertion loss and attenuation caused by the cable's greater resistance (skin effect) \cite{cable-loss}. 

As a result, DAC cable lengths are restricted: for a \mbox{200~Gb/s} IB network, HDR cables are limited to a maximum of two meters, while \mbox{100~Gb/s} EDR DAC cables can extend up to three meters. For connections at \mbox{200~Gb/s} and above, DACs can only link the TOR switch to servers within the rack or adjacent racks, particularly when wider racks (over 60~cm) are used, and cables are routed above or below the racks. Connectivity to the core IB switches is achieved via ten 200~Gb/s HDR links, optimized through a patch panel that minimizes the length of patch cables required for connecting each TOR. To streamline control of the cooling system, IT loads were distributed as evenly as possible across the containers.

To optimize cooling efficiency, the building blocks (sets of three consecutive racks containing servers connected to the same IB switch) within the containers are positioned directly beneath one of the cooling units. However, this arrangement introduced gaps between the building blocks, necessitating the rearrangement of some racks during the farm expansion and the addition of new building blocks (details on the farm extension are in Section~\ref{epn-op}).

In addition to compute servers, the containers house the majority of the infrastructure equipment required to operate all services. This includes core network switches, as well as connectivity components linking the EPN farm to the readout farm (FLP) and CERN IT storage systems (EOS), as illustrated in Fig.~\ref{fig:dataflow}.

\section{Operation of the O$^2$ EPN farm}\label{epn-op}

In 2022, data~flow stress tests were conducted using proton-proton (\mbox{pp}) collisions to simulate high-multiplicity track loads similar to those in \mbox{Pb--Pb} events. Although the track characteristics in high-rate pp collisions differ from \mbox{Pb--Pb} collisions, such tests aimed to replicate a comparable load on the detectors. Notably, \mbox{Pb--Pb} events exhibit higher charge fluctuation compared to \mbox{pp} collisions, which can generate more detector hits due to secondary particles and interactions with detector material. Initial tests revealed that the available computing resources might be insufficient for real-time data processing, prompting the addition of 30 MI-50 worker nodes to the computing farm to extend processing capacity.

Toward the end of 2022, further analysis of the \mbox{pp} data uncovered a 30\% increase in TPC data size, characterized by larger and more numerous clusters than anticipated from Monte Carlo simulations. To maintain the compute margin required to handle the foreseen 50 kHz \mbox{Pb--Pb} collision rate, an expansion of the EPN farm was approved, with plans to add 70 more powerful MI-100 servers in 2023. However, in 2022, the geopolitical situation in 2022 led to energy shortages, resulting in an early shutdown of CERN's accelerator operations. Reduced run time, coupled with commissioning delays, postponed the first high-rate \mbox{Pb--Pb} run to late 2023.

Since the 2022 high-rate \mbox{Pb--Pb} run was canceled, EPN synchronous processing validation was performed using \mbox{pp} collisions. Calculations indicated that, under a 50~kHz \mbox{Pb--Pb} collision rate, the EPN farm would need to manage at least 800~GB/s of data post-TPC ZS. The final TPC firmware was still unavailable and the validation was conducted using an intermediate firmware version. A rate scan was conducted to determine the performance limits of the synchronous processing chain. Results demonstrated that online processing machinery could sustain data rates up to 1.24~TB/s, nearly double the original nominal design rate of 600~GB/s planned for \mbox{Pb--Pb} collisions. 

It is important to note that the 2 MHz \mbox{pp} collisions dataset is not intended for physics analysis; its sole purpose was to validate the mechanics of synchronous processing. Furthermore, the zero suppressed data format used in 2022 differs from the final format, meaning the rates are not directly comparable with those achieved during the \mbox{Pb--Pb} data in 2023 (800 GB/s peak).

Additional adjustments were performed on the basic data structure for continuous readout, the time frame (TF).
In this approach, processing is not triggered by specific detector signal patterns; instead, all data is read out and stored within a predefined time window. The TF length is adjustable, typically set as a multiple of one LHC orbit, and the entire TF is processed in a single operation, necessitating that it fits within the GPU’s memory. As a result, GPU memory reuse across processing steps became essential. However, due to variations in event centrality and luminosity, the number of TPC hits fluctuates slightly, requiring a safety margin in memory allocation. For most cases, a 24 GB GPU is sufficient, with only 0.1\% of TFs exceeding this capacity. Since the current EPN farm is equipped with 32 GB GPUs, memory limitations are not a concern. 

During the initial 2022 validations, the TF length was configured to 128 LHC orbits (11.5 ms), with tests confirming stable performance of the EPN GPU setup up to 256 orbits. However, further analysis revealed advantages in reducing the TF length to 32 LHC orbits (2.8 ms). This shorter TF length boosted compute performance by 10–14\% on MI-50 GPUs and, due to a reduced memory footprint, also enabled more efficient asynchronous processing on CPU-only remote GRID sites with lower performance.

The EPN software module, data distribution (DD), manages the generation of sub time~frames (sTFs), which are partial time frames containing data from only one detector, directly at the readout farm level. This module also handles the scheduling and aggregation of STFs into complete TFs at the EPN level. Each EPN node receives and processes a full TF in sequence, combining data from all detectors, but only over the span of a single TF.

Calibration tasks for the detectors may run either on the readout nodes or on the EPNs, depending on the specific calibration type. For instance, online calibrations are confined to the EPNs and run on dedicated, CPU-only nodes. In general, any calibration task that requires access to global data and operates on entire TFs is executed on the EPN farm, whereas detector-specific calibrations that do not require such global information may be processed locally on the readout nodes.

\subsection{Compact encoding}\label{ctf}

The TF data from each sub-detector is independently reduced and compressed using algorithms specific to the sub-systems. Lossy methods remove or replace data for size reduction while lossless techniques restate the information in a more space efficient form. For each sub-detector, the resulting data is a flat structure of integer arrays. Each of the arrays is then compressed individually via a custom compression scheme based on range Asymmetric Numeral Systems (rANS \cite{ans-coding}) entropy coding.
Entropy coding leverages the probability distribution of the source data to transform each source symbol into part of a bit stream. Frequent symbols contribute fewer bits, while rarer symbols require more, optimizing the expected bit-stream length. The lower bound of this expected length is determined by the entropy rate, a theoretical limit derived from information theory. The achievable compression is therefore inherently tied to to the distribution of its source symbols. For (compressed time frame) CTF data, the entropy limit suggests a maximum compression ratio of factor 2$\times$--3$\times$.
Thanks to its capability to represent skewed probability distributions of 32-bit symbols with high fidelity, rANS achieves compression of TF data close to the entropy limit with negligible overhead. When compared to Huffman \cite{huff} coding, CTF sizes are, on average, 3\% smaller. Relative to traditional compression libraries such as \verb|gzip| or \verb|zlib|, rANS achieves up to 15\% smaller sizes, as these libraries are less efficient with alphabets larger than 8 bits per symbol.
The ALICE implementation of rANS is vectorized using AVX2 (Advanced Vector Extensions 2) \cite{avx2} instructions, enabling up to 16 encoders to work in parallel on a single stream. This implementation achieves compression speeds of up to 10$^9$ symbols per second, equivalent to 3200 MB/s for 32-bit symbols, a 2$\times$ speedup state-of-the-art CPU implementations. The metadata which is required for decoding the bit-stream is also efficiently compressed to reduce storage overhead. These performance gains free up computational resources, making it feasible to compute symbol distributions dynamically for each TF, rather than relying on pre-trained distributions. This guarantees that CTFs are consistently compressed with the minimum overhead relative to their entropy.

The implementation of CTFs is crucial for \mbox{ALICE}'s continuous data-taking operations, which face the challenge of managing the vast amounts of high-rate \mbox{pp} data (up to 1~MHz) while preparing for the \mbox{Pb--Pb} datasets. Key statistics of the collected data for different collision systems from 2022 to present are given below: 
\begin{itemize}
    \item 2022: 52~PB of \mbox{pp} data collected (no \mbox{Pb--Pb}   running)
    \item 2023: 38~PB of \mbox{pp} data and 42~PB of \mbox{Pb--Pb}  data collected
    \item 2024: 180 PB of \mbox{pp} data and 39~PB of \mbox{Pb--Pb} data collected.
\end{itemize}

To address this challenge, a strategy was developed that involves selecting offline only the most relevant events (in terms of beam bunch crossing) and skimming the CTFs to retain only between 3-4.5\% of the original data on disk (see Section \ref{computing-infra}). This approach is indispensable to avoid filling the disk buffers at CERN IT and to prevent interruptions in data taking, ensuring the sustainability of \mbox{ALICE}'s operations.

\subsection{Processing and calibration}\label{proc-calib}

\begin{figure}
    \centering
    \includegraphics[width=0.75\linewidth]{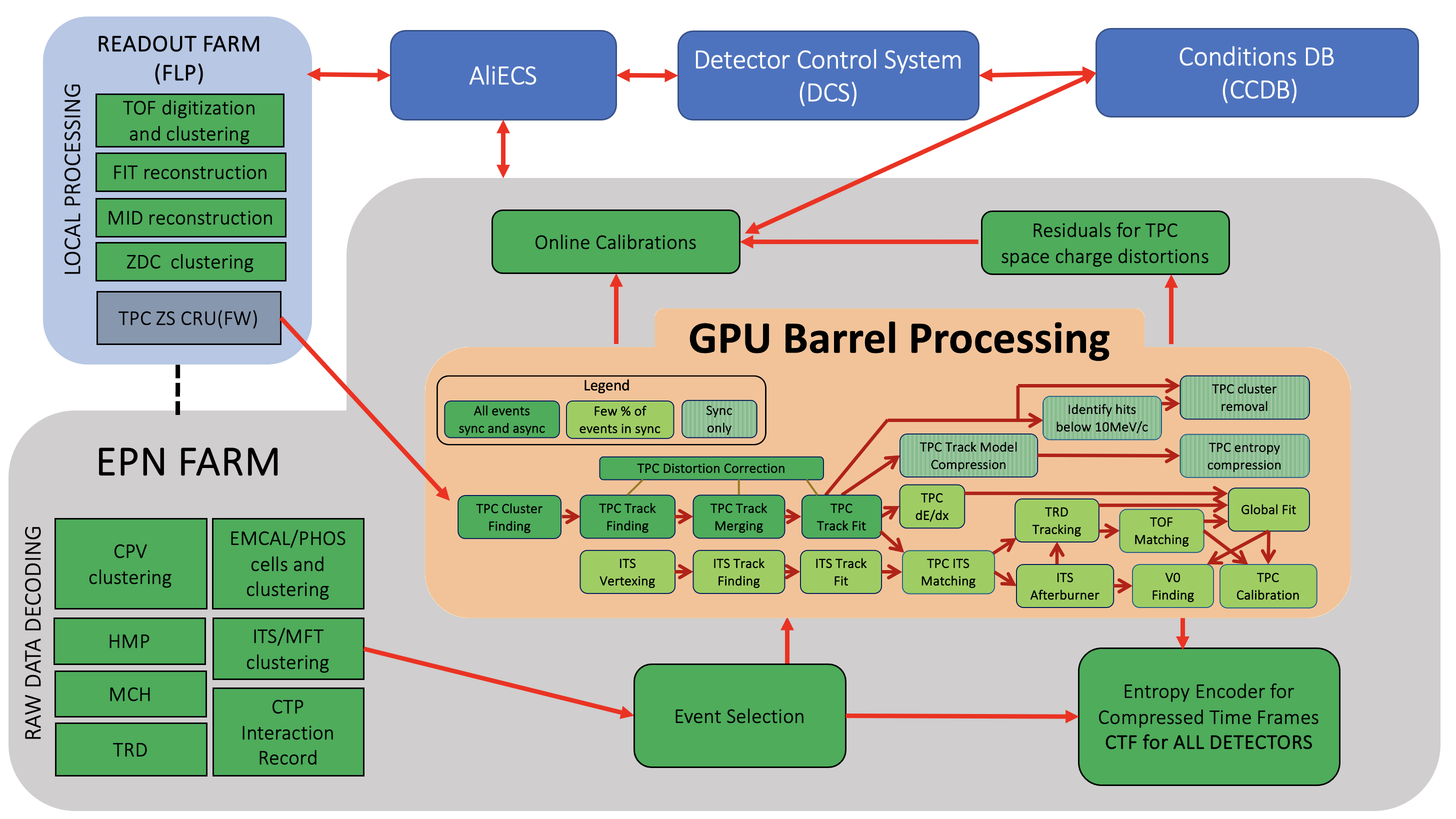}
    \caption{\small Block diagram of the O\(^2\) synchronous processing and online calibrations. Details are given in Section \ref{proc-calib}. }
    \label{fig:sync-block}
\end{figure}

The \mbox{Run 3} computing model performs data reconstruction in while data taking is ongoing (synchronous processing). In order to accomplish this task the first pass of detector calibration must also happen online in contrast with \mbox{Run 1} and \mbox{Run 2}, were the calibration would start days after the end of data takings. The full overview of the synchronous process is illustrated in detail in Fig. \ref{fig:sync-block}.
Raw STFs originating from the detector front-end electronics undergo initial local processing on the FLP farm nodes (left block) before being transferred to the EPN farm via RDMA (remote direct memory access) over IB. 

In the EPN farm, additional local processing, including raw data decoding (lower-left block), is carried out and the full TFs are built.
During synchronous operation the TPC reconstruction fully loads the GPUs with the farm providing 90\% of its compute performance via GPUs, (central block). Online calibrations are performed on dedicated CPU-only nodes within the EPN farm. During data taking, the most compute-intensive task is the TPC space-charge distortion evaluation, which requires matching and refitting of ITS, TPC, TRD, and TOF tracks, and therefore requires global track reconstruction for several detectors. At the increased \mbox{Run 3} interaction rate, processing of the order of 1\% of the events is sufficient to carry out the first calibration pass. Online calibrations also depend on the detector physical states which are available into the condition and calibration data base (CCDB, top-right block) used in \mbox{Run~3} to store calibration and alignment data. Finally, event selection is applied, and the resulting CTFs are transferred to permanent storage.

During the asynchronous phase the relative contribution of the central barrel (TPC) processing to the overall workload is much smaller so the GPU idle times are higher and processing is mostly CPU-limited. To leverage the full potential of the GPUs, also the non-TPC part of central-barrel asynchronous reconstruction software will required an implementation with native GPU support. Currently, around 60\% of the asynchronous workload can run on a GPU, yielding roughly a speedup factor of 2.5 on the EPN farm, compared to CPU-only processing (\cite{chep23}).

Once the remaining central barrel tracking software is fully adapted for GPU processing (with the primary bottleneck currently stemming from the ITS and TRD, which still operate in a single-threaded manner on CPUs), it is estimated that up to 80\% of the reconstruction workload will be executed on GPUs, even during the asynchronous phase.

With respect to synchronous processing, asynchronous processing includes the reconstruction of data from all detectors, and all events instead of only a subset; also physics-analysis ready objects produced from asynchronous processing are then made available on the GRID.
Therefore, the processing workload for detectors other then the TPC is significantly higher in the asynchronous phase since TPC clustering and the data compression are not necessary and the tracking runs on a smaller input data set since a subset of the detector hits were already removed during the data compression.

Asynchronous reconstruction performs efficiently on the EPN farm, leveraging its GPU computing capabilities. However, the EPN farm alone is insufficient to process the entire dataset generated during ALICE operations and a substantial portion of the asynchronous reconstruction workload is offloaded to remote CPU-only GRID sites. The workload distribution between synchronous and asynchronous reconstructions on the EPN farm is dynamically managed based on \mbox{ALICE}'s operational mode. 

During periods of \mbox{Pb--Pb} collisions, nearly all EPN resources are allocated to synchronous reconstruction. Conversely, during LHC shutdowns (including technical stops, winter breaks, and long shutdowns), the majority of the EPN farm is utilized for asynchronous reconstruction. For \mbox{pp} collision periods, ALICE operates at lower interaction rates compared to \mbox{Pb--Pb}, approximately one-third of the EPN farm is dedicated to data-taking activities, with the remaining capacity allocated to asynchronous processing tasks.

\subsection{Performance of the O$^2$ EPN system}\label{epn-perf}

After the mechanics of synchronous processing were fully validated with \mbox{pp} collisions, a brief period of \mbox{Pb--Pb} collisions at top center-of-mass energy (5.36~TeV) but with low beam intensity occurred at the end of 2022. Due to the very low interaction rates, the data input rate to the EPN farm was approximately 96 GB/s. Despite these modest rates, collecting the first \mbox{Pb--Pb} data at the new center-of-mass energy provided an opportunity to validate the synchronous processing workflow in terms of data quality, laying the groundwork for future operation at higher interaction rates.

During 2023 and more extensively in 2024, the O$^2$ synchronous processing on the EPN farm was fully deployed and successfully utilized to collect high-luminosity \mbox{Pb--Pb} data. Notably, in 2024, the LHC achieved higher bunch intensities and sustained the 50 kHz interaction rate for longer periods (30–40 minutes) before luminosity burn-off began to take effect.

\begin{figure}
    \centering
    \includegraphics[width=0.70\linewidth]{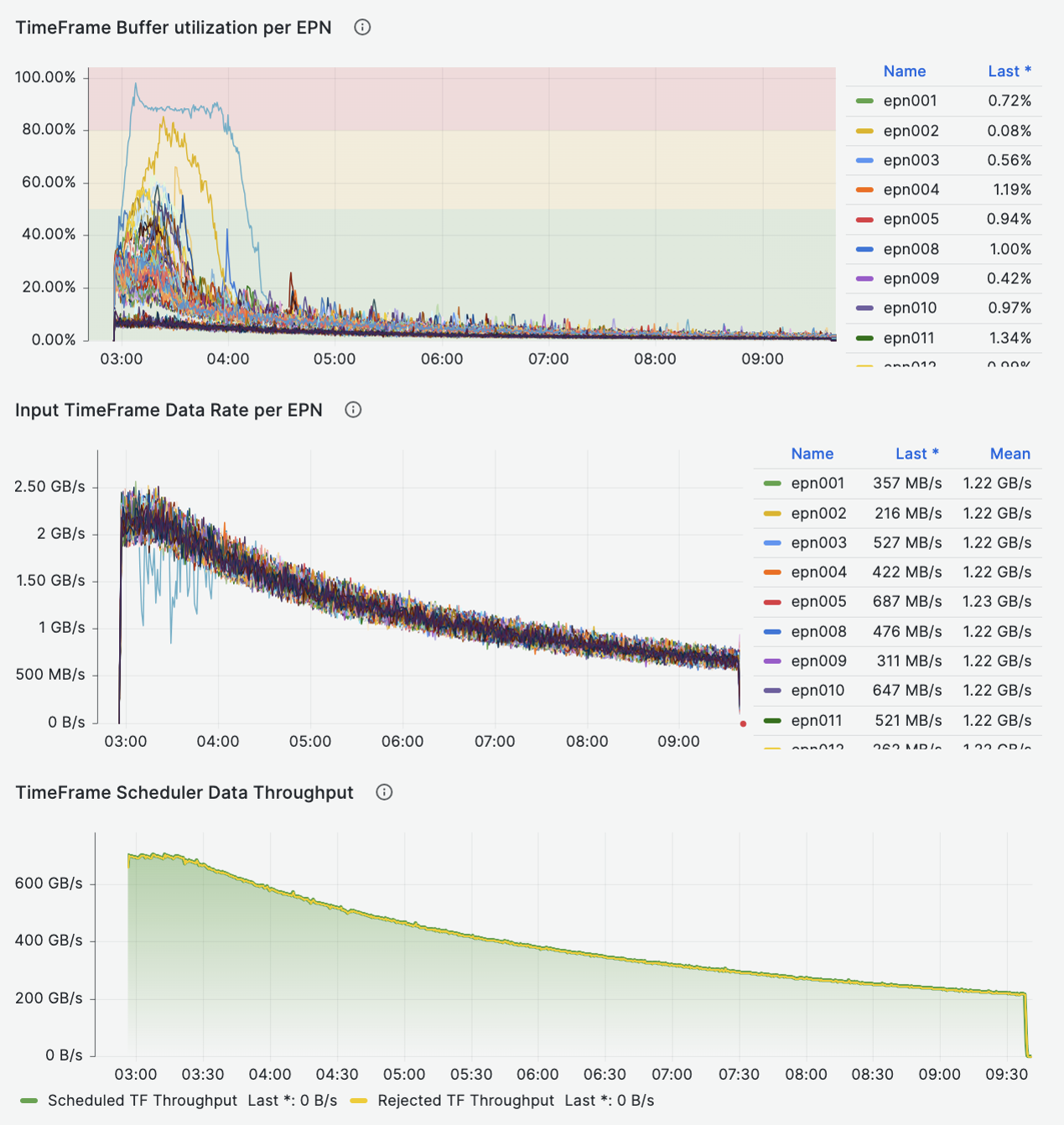}
    \caption{\small O\(^2\) EPN synchronous processing performance during the 50 kHz (sustained) \mbox{Pb--Pb} 2024 running. Details are in Section \ref{epn-perf}.}
    \label{fig:pbpb2024}
\end{figure}

A few performance metrics of the EPN farm are depicted in Fig. \ref{fig:pbpb2024}. The top graph illustrates memory utilization across the farm during data-taking. A clear performance differential is visible between the slower EPN MI-50 nodes (equipped with 2 AMD EPYC 7452 32-core CPUs and 512 GB of DDR4 3200 MHz RAM) and the faster MI-100 nodes (featuring 2 AMD EPYC 7552 48-core CPUs and 1 TB of DDR4 3200 MHz RAM). The slower nodes are equipped with less memory and require more time to process the assigned raw TFs, resulting in higher buffers utilization due to the accumulation of raw TFs waiting to be processed, especially at peak luminosities. In contrast, the faster nodes maintain nearly constant memory utilization under similar conditions. The increasing buffer utilization does not represent an issue for the processing: the initial peak is usually reabsorbed by the round-robin load balancing in DD. In addition, even if the buffers of the slow nodes would saturate the faster nodes would start to pick up more TF for processing, re-balancing the system resource usage. 

The center graph shows the input TF rate per EPN node, while the bottom distribution presents the aggregated rate from the TF Scheduler (per node). These plots reflect the natural decay of luminosity. Importantly, in the bottom plot, the rejected TF rate is zero, indicating that the system handles raw input rates effectively, with no data loss during synchronous reconstruction.

\section{Conclusions}
The \mbox{ALICE} experiment has been a pioneer in utilizing GPUs for online data reconstruction and compression in high-energy physics for over a decade. The current \mbox{ALICE} setup for \mbox{Run~3} and beyond leverages server-grade GPUs to accelerate both synchronous and asynchronous processing. 

Synchronous processing occurs during data-taking and involves tasks such as online calibrations, tracking, and efficient entropy-based lossless compression. The  \mbox{ALICE} TPC, the primary contributor to the data size, utilizes 99\% of the EPN farm's GPU compute power during this phase. 

In contrast, approximately 60\% of asynchronous processing for 650 kHz \mbox{pp} collisions is \mbox{GPU-accelerated}. This limitation arises because reconstruction for some \mbox{ALICE} detectors has not yet been ported to GPUs. Efforts are underway to increase GPU usage, aiming for 80\% GPU-accelerated code coverage for full barrel tracking.

The allocation of resources between synchronous and asynchronous reconstruction is dynamically managed, adapting to \mbox{ALICE}'s operational needs, interaction rates, and collision system types. 

GPUs offer exceptional processing efficiency, delivering high compute performance and data quality at a lower cost compared to CPU-only processing. Their effectiveness, compactness, and favorable cost-benefit ratio are increasingly drawing interest from the high-energy physics community, including other LHC experiments \cite{courier}. 

The \mbox{ALICE} EPN IT infrastructure hosts 350 servers equipped with 2800 GPUs and employs energy-efficient techniques, such as adiabatic cooling, to reduce its carbon footprint and enhance power usage effectiveness. 

In conclusion, GPU-based HPC computing coupled with an energetically efficient data center infrastructure appear to be the most economically viable and low environmental impact solution for meeting the computational demands of high-energy physics experiments.

\onecolumn

\section*{Acknowledgments}
The authors wish to thank the Bundesministerium für Bildung und Forschung (BMBF) of Germany for supporting the EPN project.

\section*{Figure captions}

\newpage

\bibliographystyle{unsrt}
\bibliography{references.bib}

\clearpage

\end{document}